\documentclass[twocolumn,showpacs,preprintnumbers,amsmath,amssymb,APSl,prd,nofootinbib,superscriptaddress]{revtex4-1}

\usepackage{dcolumn}
\usepackage{bm}
\usepackage{ifpdf}
\usepackage{hyperref}
\usepackage{dcolumn}
\usepackage{bm}
\usepackage[spanish,english]{babel}
\usepackage{amsfonts}
\usepackage{amssymb}
\usepackage{graphicx}
\usepackage{color}
\usepackage[latin1]{inputenc}

\newcommand{\be}{\begin{equation}}
\newcommand{\en}{\end{equation}}
\newcommand{\bea}{\begin{eqnarray}}
\newcommand{\ena}{\end{eqnarray}}

\begin{document}

\title{Wormholes and nonsingular space-times in Palatini $f(R)$ gravity}

\author{Cosimo Bambi} \email{bambi@fudan.edu.cn}
\affiliation{Center for Field Theory and Particle Physics and Department of Physics, Fudan University, 220 Handan Road,
200433 Shanghai, China.}
\affiliation{Theoretical Astrophysics, Eberhard-Karls Universit\"at
T\"ubingen, 72076 T\"ubingen, Germany.}
\author{Alejandro Cardenas-Avendano} \email{alejandro.cardenasa@konradlorenz.edu.co}
\affiliation{Programa de Matem\'atica, Fundaci\'on Universitaria Konrad Lorenz, 110231 Bogot\'a, Colombia.}
\affiliation{Observatorio Astron\'omico Nacional de Colombia, Universidad Nacional de Colombia, 111321 Bogot\'a,
Colombia.}
\author{Gonzalo J. Olmo} \email{gonzalo.olmo@csic.es}
\affiliation{Departamento de F\'{i}sica Te\'{o}rica and IFIC, Centro Mixto Universidad de
Valencia - CSIC. Universidad de Valencia, Burjassot-46100, Valencia, Spain.}
\affiliation{Departamento de F\'isica, Universidade Federal da
Para\'\i ba, 58051-900 Jo\~ao Pessoa, Para\'\i ba, Brazil.}
\author{D. Rubiera-Garcia} \email{drgarcia@fc.ul.pt}
\affiliation{Instituto de Astrof\'isica e Ci\^encias do Espa\c{c}o, Universidade de Lisboa, Faculdade de Ci\^encias, Campo Grande, PT1749-016 Lisboa, Portugal}
\affiliation{Center for Field Theory and Particle Physics and Department of Physics, Fudan University, 220 Handan Road,
200433 Shanghai, China.}

\date{\today}

\begin{abstract}
We reconsider the problem of $f(R)$ theories of gravity coupled to Born-Infeld theory of electrodynamics formulated in
a Palatini approach, where metric and connection are independent fields. By studying electrovacuum
configurations in a static and spherically symmetric space-time, we find solutions which reduce to their
Reissner-Nordstr\"om counterparts at large distances but undergo important non-perturbative modifications close to
the center. Our new analysis reveals that the point-like singularity is replaced by a finite-size
wormhole structure, which provides a geodesically complete and thus nonsingular space-time, despite the existence of curvature divergences at the wormhole throat. Implications of these results, in particular for the cosmic censorship conjecture, are
discussed.
\end{abstract}


\maketitle

\section{Introduction}

The quest for regular solutions has historically been a powerful driving force in Physics. A well known example is
given by the problem of the divergence of Coulomb's field in classical electrodynamics. Consideration of this
problem gave rise to the development of classical models of the electromagnetic field, most remarkably the one
introduced by Born and Infeld \cite{BI34}. In such model, at far distances Coulomb's field is recovered, while
deviations close to the center yield a finite total energy, though singularities are still present under the form of delta-like sources at $r=0$. As is well known, the full resolution of this problem was only achieved within the framework of quantum theory.

Space-time singularities are different beasts. As opposed to singularities of the fields living on a fixed space-time
background, these are singularities of the background itself. Indeed, because
they represent a breakdown of the geometry, this even prevents to speak of singularities as some
troublesome event occurring in a specific location of the space-time, since the absence of a well defined geometry
makes one
not able to speak on such a ``location" (see for instance the discussion presented in Ref.\cite{Senovilla}). This
difficulty was explicitly acknowledged during the development of the singularity theorems \cite{Theorems}. To overcome
it, one focuses instead on
the idea of geodesic completeness and singularities are thus attached to the existence of geodesic curves terminating
on a finite length of the affine parameter \cite{Geroch}. In this way, under the requirements of the existence of
trapped surfaces, global hyperbolicity and the fulfillment of the null congruence condition (which is equivalent to the
energy conditions, provided that the Einstein
equations of General Relativity (GR) hold), the singularity theorems tell us that space-time singularities are
unavoidable \cite{Hawking}.

When a geodesic curve cannot be arbitrarily extended, one is tempted to assume that this happened due to the existence
of some kind of pathology in the structure of space-time, like the blow up of curvature scalars for instance. Although
this is not
what the theorems on singularities say, there is a widespread identification between space-time singularities and
curvature divergences in the literature, which has shaped numerous attempts to overcome the singularity problem in the
context of GR, by constructing ad hoc classical geometries where the curvature scalars are bounded (see
e.g., Ref. \cite{Ansoldi} for a review). This has been highly influenced by the success of Bardeen's solution
\cite{Bardeen}, where a magnetic monopole, solution of a non-linear electrodynamics model \cite{ABG}, gets rid of
curvature divergences (and becomes geodesically complete) by introducing a de Sitter core at the center
\cite{Dymnikova}. This gives a great relevance to classical models like the Born-Infeld one, which consequently have
been widely studied \cite{BI-gravity}, with the result that curvature divergences can be softened, but not completely
got rid of. The latter requires instead more exotic models violating the energy conditions or which are physically
troublesome \cite{AB-regular,Bronnikov}. Nonetheless, let us stress that geodesic completeness is regarded in the
literature as the most fundamental criteria to discuss space-time singularities\footnote{This is because some geodesics,
being interpreted as the existence of physical observers, should exist and be defined everywhere, while the presence of
large deformations suffered by those observers becomes secondary as long as geodesics exist.} above the intuitive
criterion of curvature scalars \cite{Theorems,Hawking,Wald}, and that no identification exists between the former and
the latter.

A different viewpoint on this issue, is to assume that the dynamic at very short scales is not fully captured by GR.
Space-time singularities are thus viewed as an artifact resulting from the inadequacy of GR to deal with extreme
scenarios, like the innermost region of black holes.  One of the simplest extensions of GR is that of $f(R)$ gravity,
which has been extensively studied in the literature (see for instance Ref.\cite{fR-review} for some reviews and also \cite{fR-papers} for some recent work). In
this sense, in Ref.\cite{or11} two of us considered an $f(R)=R-\gamma R^2$ model, formulated in the Palatini approach, where the metric and connection are not constrained a priori, but instead determined by independent variations of the action \cite{olmo-review}. Such an approach is supported by the effective geometries arising from the study of the continuum limit of crystalline systems with defects on their microstructure \cite{ssp,lor15}, where different types of defects require different geometric structures. The Palatini formulation hinges on the fact that metric-affine geometric scenarios might be relevant in gravitational physics too.

The new gravitational dynamics present on $f(R)$ theories can only be excited when coupled to a matter source whose energy-momentum tensor is non-traceless.
This prevents using Maxwell electrodynamics to obtain modifications with respect to GR solutions. For this reason, in Ref.\cite{or11} the Born-Infeld (BI) nonlinear electromagnetic theory was used, which allowed the authors to compare the solutions with those obtained in GR in static and  spherically symmetric space-times. That work identified the existence of solutions which reduce to the BI counterpart of GR at large distances, but undergo
important modifications near the center with two main features, i.e., the existence of a core at a distance $r=r_c \neq 0$, and the softening of curvature
divergences, from the $\sim 1/r^4$ divergence of the BI-GR case [and the $\sim 1/r^8$ of the Reissner-Nordstr\"om (RN) solution], down to $\sim 1/(r-r_c)^2$.

Supported on a series of results obtained in the last few years regarding other Palatini theories \cite{or,ors}, the aim of this paper is to revisit the results of Ref.\cite{or11} (which are summarized in Sec. \ref{sec:II}, for the reader's convenience and completeness) and identify two new important aspects previously unnoticed. First, in Sec. \ref{sec:III} we show that the core at $r=r_c$ actually represents the throat of a wormhole. Second, in Sec. \ref{sec:IV} we show that the presence of the wormhole yields a geodesically complete space-time for both null, time-like and space-like geodesics, no matter the behavior of the curvature scalars. The geodesic completeness of these solutions puts forward that this $f(R)$ theory with a matter source satisfying the energy conditions is yielding nonsingular space-times, which is in sharp contrast with the RN and BI solutions of GR, and the standard approaches to the resolution of space-time singularities typically found in the literature. The conclusions are presented in Sec. \ref{sec:V}.

\section{Revisiting Palatini $f(R)$ gravity with a Born-Infeld field} \label{sec:II}

In this section we shall revisit the main results developed in Ref. \cite{or11}, corresponding to an action
$S=S_G+S_{BI}$, where $S_G$ is the gravitational sector

\begin{equation}
S_{G}=\frac{1}{2\kappa^2} \int d^4x \sqrt{-g} f(R),
\end{equation}
where $\kappa^2=8\pi G $ (in units $c=1$), $g$ is the determinant of the space-time metric $g_{\mu\nu}$, which is
independent of the affine connection $\Gamma \equiv \Gamma_{\mu\nu}^{\rho}$ (Palatini or metric-affine approach
\cite{olmo-review}), $R=g^{\mu\nu} R_{\mu\nu}(\Gamma)$ is the curvature scalar constructed with the Ricci tensor
$R_{\mu\nu}(\Gamma)$, which only depends on the affine connection $\Gamma$, and $f(R)=R-\gamma R^2$ is the particular
model we will be interested in this work, with $\gamma$ a small parameter with dimensions of length squared. In their
more standard metric formulation (i.e., a connection compatible with the metric \emph{a priori}), these types of quadratic
$f(R)$ models have attracted a lot of attention in the context of inflation \cite{fR-inflation}. For the purposes of
this work we shall assume $\gamma >0$. The matter sector is described by Born-Infeld theory of electrodynamics, $S_{BI}
= (8\pi)^{-1}\int d^4x \sqrt{-g} \varphi_{BI}(X,Y)$, with

\begin{equation} \label{eq:BI}
\varphi_{BI}(X)= 2\beta^2 \left(1-\sqrt{1-\frac{X}{\beta^2} - \frac{Y^2}{4\beta^4} } \right),
\end{equation}
where $X=-\frac{1}{2} F_{\mu\nu}F^{\mu\nu}$ and $Y=-\frac{1}{2} F_{\mu\nu}F^{*\mu\nu}$ are the field invariants which
can be constructed with the Maxwell field strength tensor $F_{\mu\nu}=\partial_{\mu}A_{\nu} - \partial_{\nu}A_{\mu}$
and its dual $F^{*\mu\nu}=\frac{1}{2} \epsilon^{\mu\nu\alpha\beta}F_{\alpha\beta}$. Note that in the limit $\beta
\rightarrow \infty$ the Maxwell Lagrangian, $\varphi(X)=X$, is recovered from (\ref{eq:BI}).

By introducing an auxiliary metric $h_{\mu\nu}= f_R g_{\mu\nu}$, which is conformally related with $g_{\mu\nu}$, the field equations in Palatini $f(R)$ gravity can be cast in Einstein-like form \cite{or11}
\begin{equation} \label{eq:equations}
{R_\mu}^{\nu}(h)=\frac{1}{f_R^2} \left(\frac{f}{2} {\delta_\mu}^{\nu} + \kappa^2 {T_\mu}^{\nu} \right),
\end{equation}
We recall, see \cite{or11}  for details, that $f_R\equiv df/dR$ is a function of the trace  $T$  of the energy-momentum tensor [in the particular case $f(R)=R-\gamma R^2$, one has that $R=-\kappa^2 T$, which is the same relation as in GR], which means that the right-hand-side of the field equations (\ref{eq:equations}) is just a function of the matter. This implies that (\ref{eq:equations}) represents  a system of second-order field equations for the metric $h_{\mu\nu}$, which is in sharp contrast with the generic fourth-order field equations that one finds in the metric formulation of these theories \cite{fR-review}.

From the generic expression of the energy-momentum tensor for non-linear electrodynamics
\begin{equation}
{T_\mu}^{\nu}=\frac{1}{8\pi}\text{diag}[\varphi-2X\varphi_X, \varphi-2X\varphi_X, \varphi,\varphi]
\end{equation}
one verifies that the trace, $T=\frac{1}{2\pi} [\varphi - X\varphi_X]$, is non-vanishing for the case of Born-Infeld theory. This way, one is able to provide modified dynamics to the field equations (\ref{eq:equations})  and, therefore, to $g_{\mu\nu}$. Note that in vacuum, ${T_\mu}^{\nu}=0$, one has $g_{\mu\nu}=h_{\mu\nu}$ and the field equations (\ref{eq:equations}) reduce to those of GR with a cosmological constant and, therefore, the theory is free of ghosts. Let us point out that both the second-order character of the field equations and the absence of ghost is a generic characteristic of the Palatini formulation of modified gravity, which holds for models including Ricci-squared terms \cite{or}, in Born-Infeld gravity actions \cite{ors,oor}, and in higher-dimensions \cite{higher-dim}.

Solving the above equations for $h_{\mu\nu}$ and using the conformal transformation at the end, a solution for
$g_{\mu\nu}$ in a (electro-)static, spherically symmetric space-time with $X=F_{tr}^2, Y=0$ can be obtained in exact
form \cite{or11} . To simplify the notation, we introduce a length scale for the charge $r_q^2=\kappa^2 q^2/(4\pi)$ and
another for the Born-Infeld nonlinearities $l_{\beta}^2=1/(\kappa^2 \beta^2)$, and define a new dimensionless coordinate $z\equiv
r/r_c$, where $r_c\equiv (4\pi)^{1/4} \sqrt{r_q l_{\beta}}$. Using this notation, the line
element can be conveniently written as

\begin{equation} \label{eq:linelement}
ds^2= \frac{1}{f_R} \left(-A(x)dt^2 + \frac{1}{A(x)} dx^2 + x^2 d\Omega^2 \right),
\end{equation}
where the function $A(x)$ can be written by introducing the standard mass ansatz $A(x)=1-2M(x)/x$, which in this case
is given by
\begin{equation} \label{eq:Afunction}
A(x)=1-\frac{1+\delta_1 G(z)}{\delta_2 z f_R^{1/2}} \ ,
\end{equation}
where the function $G(z)$ satisfies
\begin{equation} \label{eq:Gz}
\frac{dG}{dz}= \frac{z^2}{4f_R^{3/2}} \left(\tilde{f} + \frac{\tilde{\varphi}}{4\pi} \right) \left(f_R + \frac{z}{2}
f_{R,z} \right) \ ,
\end{equation}
and we have the important relation
\begin{equation}\label{eq:z(x)}
x^2=f_R z^2 \ ,
\end{equation}
which follows from the conformal relation between $h_{\mu\nu}$ and $g_{\mu\nu}$ and allows us to obtain $z=z(x)$. Deriving Eq.(\ref{eq:z(x)}) we also obtain the equation

\begin{equation} \label{eq:dxdz}
\frac{dx}{dz}=f_R^{1/2} \left(1+\frac{zf_{R,z}}{2f_R} \right)
\end{equation}
which shall be useful later.

The function $G(z)$ gets contributions from two sides: the $f(R)$ gravity

\begin{eqnarray} \label{eq:f}
\tilde{f} &\equiv& l_{\beta}^2 f= \frac{\eta(z)}{2\pi} \left(1-\frac{\alpha}{2} \eta(z) \right)  \\
f_R&=&1-\alpha \eta(z),
\end{eqnarray}
where $\alpha \equiv \gamma/(2\pi l_{\beta}^2)$, and the Born-Infeld model

\begin{eqnarray} \label{eq:v}
\tilde{\varphi}(z)&=&2\left(1-\frac{1}{\sqrt{1+1/z^4}} \right) \\
\eta(z)&=& \frac{\left(z^2-\sqrt{1+z^4}\right)^2}{z^2\sqrt{1+z^4}}.
\end{eqnarray}
In this way, the problem is characterized by three parameters, namely

\begin{equation}
\delta_1=2(4\pi)^{3/4} \frac{r_q}{r_S} \sqrt{\frac{r_q}{l_{\beta}}} \ ; \  \delta_2=\frac{r_c}{r_S}
\end{equation}
(where $r_S \equiv 2M_0$ is Schwarzschild radius) and $\alpha$. In the limit $\alpha \rightarrow 0$
we have $f_R \rightarrow 1$ and thus we recover the solutions of Born-Infeld theory in GR \cite{BI-gravity}. On the other hand, the
limit $\alpha \rightarrow \infty$ corresponds to the Maxwell theory which, as was already mentioned, takes us back to
the RN space-time of GR.

\section{Wormhole structure and main features} \label{sec:III}

The space-time described by the line element (\ref{eq:linelement}) corresponds to a modification of the solution of Born-Infeld electrodynamics coupled to GR, which is recovered a few units away from the surface $r=r_c$, where large non-perturbative corrections occur. For $z \gg 1$ one obtains

\begin{equation}
g_{tt} \simeq 1 - \frac{1}{\delta_2 z} +\frac{\delta_1}{16\pi \delta_2 z^2}  + \frac{\delta_1}{320\pi \delta_2 z^6} +
\frac{\gamma}{4z^8} + O\left(\frac{1}{z^9} \right).
\end{equation}
The first three terms correspond to the RN solution with mass and charge of GR. Next we find the
Born-Infeld induced corrections (where $\gamma=\gamma(\alpha)$ is some constant) and those of $f(R)$ gravity, both of which are largely suppressed. In terms of horizons, a numerical analysis reveals \cite{or11} that the typical structure of two horizons, a single (degenerate) one, or none of the RN solution is restored, which is in
contrast with the additional structures found in the Born-Infeld-GR case, where black holes with a single (non-degenerate) horizon may arise \cite{BI-gravity}.

\begin{figure}[h]
\includegraphics[width=0.45\textwidth]{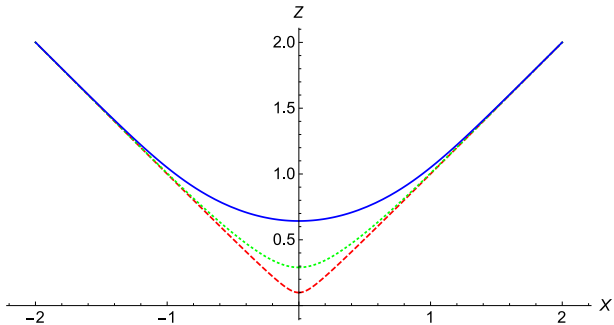}
\caption{Behavior of the radial function $z(x)$. For large $|x|$, one has $z^2 \sim x^2$ and the standard GR result. Note the bounce of the radial coordinate at $z = z_c(\alpha)$. This behavior is reminiscent of a wormhole geometry with $z=z_c(\alpha)$ the location of the wormhole throat. When $\alpha=0$ the throat closes and we recover the GR point-like singularity. The dashed (red) curve has $\alpha=10^{-2}$, the green (dotted) curve $\alpha=10^{-1}$, and the solid (blue) curve corresponds to $\alpha=1$.  \label{fig:z(x)}}
\end{figure}

To investigate the structure of the region $r \simeq r_c$ in detail, Eq. (\ref{eq:z(x)}) is very important. Given that an analytical expression for $z=z(x)$ is difficult to write, in Fig.\ref{fig:z(x)} we provide a graphic representation for different values of the parameter $\alpha$. From this plot it is apparent that $z(x)$ reaches a minimum of magnitude $z_c(\alpha)$, where
\begin{equation}
z_c(\alpha)=\frac{1+2\alpha-\sqrt{1+4\alpha}}{2\sqrt{1+4\alpha}} \ ,
\end{equation}
at the location $x=0$. This minimum in the area function denotes the existence of a wormhole \cite{Visser} and prevents the use of $z$ as a good radial coordinate across the throat.

To further understand this geometry around $z \simeq z_c$, we expand the metric functions there. For $G(z)$ we find

\begin{equation} \label{eq:Gz}
G(z)\simeq -\frac{a(\alpha)}{(z-z_c)^{1/2}} + b(\alpha) (z-z_c)^{1/2} + O(z-z_c)^{3/2},
\end{equation}
where $a(\alpha)>0$, $b(\alpha)>0$ are some constants with an involved dependence on $\alpha$. Similarly, using the expansion of $f_R$:

\begin{equation} \label{eq:fRex}
f_R \simeq c(\alpha)(z-z_c) -d(\alpha)(z-z_c)^2 + O(z-z_c)^3,
\end{equation}
where $c(\alpha)>0$, $d(\alpha)>0$ is another pair of constants. From Eq.(\ref{eq:Afunction}) one finally gets

\begin{equation} \label{eq:Aex}
A(x) \simeq \frac{\delta_1 a(\alpha)}{\delta_2 z_c c(\alpha) (z-z_c)^2}.
\end{equation}
The divergent behavior of $A(x)$ at $z=z_c(\alpha)$ (i.e., $x=0$) induces similar divergences in the curvature
invariants. For example, the Kretchsmann scalar, $K = {R_\alpha}^{\beta\gamma\delta} {R^\alpha}_{\beta\gamma\delta}$
diverges there as

\begin{equation}
K \sim \frac{1}{(z-z_c)^2}.
\end{equation}
These divergences occur on a spherical surface of area $S(z_c)=4\pi z_c^2$, and are much milder than in the case of GR: in the standard RN solution one finds a divergence $\sim 1/r^8$, which can be tamed down to $1/z^4$ in the Born-Infeld case when the model parameters are properly chosen.

Despite the appearance of curvature divergences at the surface $z=z_c(\alpha)$, the existence of the wormhole structure can be further reinforced exploring the definition of the electric charge in our theory. This can be defined through the flux integral
\begin{equation}
\Phi= \int_{S^2} *F=4\pi q,
\end{equation}
where $*F$ is Hodge dual of the field strength tensor and the integration takes over an $S^2$ sphere (or any other 2-surface) enclosing the
wormhole throat $x=0$. If we take a sphere on the $x>0$ region, the result of the integral is equal in magnitude but of opposite sign to the integral performed on the $x<0$ region due to their different orientation (the normal to the surface goes in the direction of growing $x$ when $x>0$ but in the direction of decreasing $x$ when $x<0$). The nonzero flux on each side defines a charge despite the absence of a charge density. This is a topological effect well known in the literature. In fact, this combination of a topologically non-trivial structure
with a sourceless electric field can be seen as a explicit realization of the concept of geon introduced by Wheeler \cite{Wheeler}, and
the proposed charge-without-charge mechanism \cite{MW}. It should be noted that in the GR case, i.e., when $\alpha=0$, the wormhole throat closes and the well known problems with point-like sources of GR arise \cite{Ortin}.
Let us point out that the sources problem cannot be solved for any theory of non-linear electrodynamics in the context of GR \cite{Arellano}, so the introduction of the $f(R)$ model is essential.

These wormhole solutions have further interesting properties, related to the rate of growth of the intensity of the electric flux with the wormhole throat area, i.e.,

\begin{equation}
\frac{\Phi}{4\pi r_c^2 z_c^2} = \frac{\alpha^{1/2}}{z_c^2(\alpha)} \sqrt{\frac{c^7}{8\pi \hbar G^2}}.
\end{equation}
This result is independent on the specific amounts of charge and mass present in the system, and only depends on the parameter $\alpha$ governing the deviations with respect to GR. Since as long as $\alpha \neq 0$ the electric flux is non-zero,
it follows that this theory possesses topological properties (wormhole) which are independent of its geometric aspects (curvature divergences). This way we are naturally led to study the geodesic structure in this space-time to see if its nontrivial topological structure affects its properties concerning singularities.

\section{Geodesic completeness} \label{sec:IV}

A geodesic curve $\gamma^{\mu}=x^{\mu}(\lambda)$, where $\lambda$ is the affine parameter, satisfies the equation \cite{Chandra,geofR}
\begin{equation} \label{eq:geodesic1}
\frac{d^2x^{\mu}}{d\lambda^2} + \Gamma_{\alpha\beta}^{\mu} \frac{dx^{\alpha}}{d\lambda}\frac{dx^{\beta}}{d\lambda}=0.
\end{equation}
Given that in Palatini theories the action is defined by assuming as independent the metric and the affine connection, one should clarify what connection is being used in the above equation.  In this sense, we point out that in our framework the matter sector does not couple directly to the connection, which implies that the matter energy-momentum tensor is conserved with respect to the covariant derivative defined with the Levi-Civita connection of the metric, $\nabla_{\mu}^{g} T^{\mu\nu}=0$. The physical meaning of this statement is that test particles should move on geodesics of the metric and, for this reason, here we will consider those.
In our theory, therefore, the role of the independent connection is that of modifying the field equations that generate
the metric but it does not act directly on the matter fields. This is what happens in all theories of gravity that
satisfy the Einstein Equivalence Principle, such as in scalar-tensor theories, for instance, where the additional
gravitational fields help to generate the space-time curvature but do not couple directly to the matter. Note, in
addition, that we are using a sourceless electromagnetic field whose field equations $\nabla_{\mu} (\sqrt{-g} \varphi_X
F^{\mu\nu})=0$, are insensitive to the details of the particular (symmetric) connection chosen.

The geodesic equation (\ref{eq:geodesic1}) can be alternatively written making use of the unit four vector
$u^{\mu}=dx^{\mu}/d\lambda$ as

\be \label{eq:geodesic2}
\frac{du^{\mu}}{d\lambda} + \Gamma^{\mu}_{\nu\sigma} u^{\nu} u^{\sigma}=0.
\en
Instead of solving (\ref{eq:geodesic2}) directly, we use the fact that $u_{\mu}u^{\mu}=-k$ where $k=0(1)$ for
null(timelike) geodesics. Moreover, without loss of generality, in our static spherically symmetric space-time we can rotate the coordinate system to make it
coincide with the plane $\theta=\pi/2$. We can also identify two constants of the motion, $E=A^{-1} dt/d\lambda$ and $L=r^2 d\varphi/dr$, which simplifies the analysis. For time-like geodesics $E$ carries the meaning of the total energy per unit mass and $L$ that of the angular momentum per unit mass around an axis normal
to the plane $\theta=\pi/2$. For null geodesics both $E$ and $L$ lack a proper meaning by themselves, but
we can identify $L/E$ as an impact parameter \cite{Chandra}. Under these conditions, one can write the geodesic
equation corresponding to the line element (\ref{eq:linelement}) as

\be
\frac{1}{f_R^2} \left(\frac{dx}{d\lambda} \right)^2=E^2-V(x),
\en
where

\be
V(x)=\frac{A(x)}{f_R} \left(k+\frac{L^2}{r^2(x)} \right).
\en
This equation is similar to that describing the motion of a particle in a Newtonian potential $V(x)$, and is analogous to that found in scenarios including Ricci-squared corrections to the gravitational Lagrangian \cite{ors15a}. However, in the case considered here, the fact that $f_R \rightarrow 0$ as $x \simeq 0$ restricts that interpretation to regions with large $|x|$.

Using the relation $x^2=z^2/f_R$ and Eq.(\ref{eq:dxdz}) and redefining $\tilde{\lambda} \equiv \lambda/r_c$, we can write the above equation on one of the sides of the wormhole as (where the sign $+$ ($-$) corresponds to ingoing (outgoing) geodesics)
\be \label{eq:geoeq}
\frac{d\tilde{\lambda}}{dz} = \pm \frac{f_R^{1/2}\left(1+\frac{zf_{R,z}}{2f_R}\right)}{\sqrt{E^2 f_R^2 - A(z)f_R
\left(k+\frac{L^2}{r_c^2 z^2} \right)}},
\en
which facilitates the analysis. Note that this equation holds for any $f(R)$ gravity, provided that its line element can be written under the form (\ref{eq:linelement}).

Let us first consider null geodesics with angular momentum ($k=0,L \neq 0)$, and time-like geodesics ($k=1$). We first
note that since these space-times reduce to their GR counterparts for $z \gg 1$ we can safely assume the GR geodesic
behavior there. Now, using the expansions (\ref{eq:fRex}) and (\ref{eq:Aex}) we study the relevant region around the
wormhole throat $z=z_c(\alpha)$. This way one gets the result

\begin{equation} \label{eq:geodeq}
\frac{d\tilde{\lambda}}{dz} \simeq \pm \frac{c^{1/2} z_c}{2(z-z_c)^{1/2} \sqrt{c^2E^2(z-z_c)^2-\frac{\delta_1
a}{\delta_2 z_c} \left(k+\frac{L^2}{r_c^2 z^2} \right)}}.
\end{equation}
Given the behavior of the terms within the square root, it follows that for these geodesics the square root must vanish
before the throat at $z=z_c$ is reached, regardless of the energy carried by the particle, which implies that these geodesics cannot reach the wormhole. This behavior is similar to that found in the RN solution of GR, where time-like and null geodesics with $L \neq 0$ cannot overcome
the infinite barrier generated by the central object \cite{Chandra}.

\begin{figure}[h]
\includegraphics[width=0.45\textwidth]{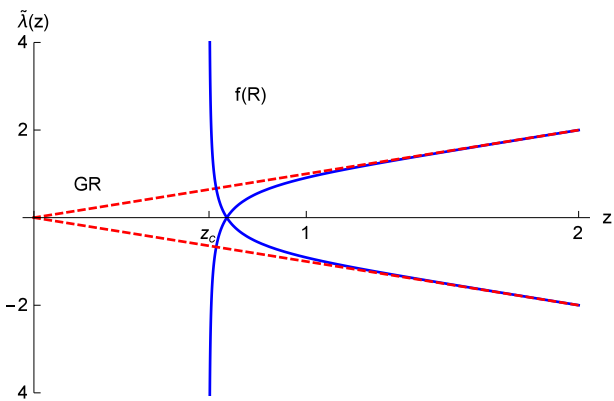}
\caption{Radial null geodesics (ingoing and outgoing), from numerical integration of (\ref{eq:nullradial}) and
$\alpha=1$ (solid blue lines). For $z \gg 1$ the geodesic behavior is the same as in GR (dashed red lines). As we
approach $z_c(\lambda)$ the deviations are manifest, and the geodesics can be extended to arbitrarily large values of
the affine parameter, in contrast to the GR case. This avoidance of singularities holds as long as $\alpha \neq 0$. \label{fig:2}}
\end{figure}

More interesting results are found for null radial geodesics ($k=0,L=0$), where the geodesic equation
(\ref{eq:geoeq}) can be written as

\be \label{eq:nullradial}
E \frac{d\tilde{\lambda}}{dz} = \pm\frac{\left(1+\frac{zf_{R,z}}{2f_R}\right)}{f_R^{1/2}}.
\en
For $z \gg 1$, this equation simply gives $E\tilde{\lambda}(z) \simeq \pm z$, which is in complete agreement with the GR prediction. Near the region $z=z_c$, using again the expansions (\ref{eq:fRex}) and (\ref{eq:Aex}), we get

\be
E\tilde{\lambda}(z) \simeq \mp \frac{z_c}{c^{1/2}(z-z_c)^{1/2}}.
\en
This is in sharp contrast with the behaviour of the standard RNsolution of GR, where the behavior
$E\tilde{\lambda}(z) = \pm z$ persists at all scales.  A numerical integration of the geodesic equation
(\ref{eq:nullradial}) for any $z$ in the $f(R)$ case has been depicted in Fig. \ref{fig:2}, and compared to the GR
behavior (dashed lines). In the GR case, radial null geodesics do reach the center, $x \rightarrow 0$, in a finite time
of the affine parameter \cite{Chandra} and the space-time is regarded as singular, due to the fact that the geometry cannot be
further extended (because $z>0$). In the GR case, therefore, there is a correspondence between geodesic incompleteness
and the divergence of curvature scalars. However, in our case, the deformation of the space-time near the region
$x=0$, due to the presence of the wormhole structure, allows null radial geodesics to be extended to arbitrarily large
values of the affine parameter, i. e.,  $\tilde{\lambda}(z) \in ]-\infty,+\infty[ $.

For completeness, let us also consider the case of space-like geodesics, i. e., $k=-1$. We note first that while the completeness of null and time-like geodesics is of direct physical significance, as they are related with physical observers and the transmission of information, the relevance of space-like geodesics being complete or not is rather unclear  (see for instance the discussion presented in Ref.\cite{Hawking}). For this reason, it is reasonable to assume that time-like and null completeness are the minimal physical conditions for a space-time to be singularity-free. Going into the point, inspection of Eq. (\ref{eq:geodeq}) reveals that the fate of these geodesics around the wormhole throat is governed by the sign of the term $(1-L^2/r_c^2)$ under the square root. In this sense, if $L^2>r_c^2$ the square root will vanish before
$r=r_c$ and thus these geodesics cannot reach the wormhole throat (like in the time-like and null cases above with $L \neq 0$), while if $L^2=r_c^2$ the behaviour is exactly the same as for radial null geodesics. When
$L^2<r_c^2$ one can integrate the geodesic equation near the wormhole throat $z=z_c$ as (modulo an integration constant) $\tilde{\lambda}(z) \simeq  \pm
((c \delta_2 z_c^3)/(a\delta_1(1-L^2/r_c^2))^{1/2} (z-z_c)^{1/2}$, which in terms of the coordinate $x$ becomes
$\tilde{\lambda}(x) \simeq \pm
((\delta_2 z_c)/(a c^2 \delta_1 (1-L^2/r_c^2))^{1/2}x$. Given that $x$ is defined on the whole real line, it is now apparent that $\tilde{\lambda}(x)$ can be smoothly extended across the wormhole throat,  which verifies that space-like geodesics in this geometry are complete as well.

The picture that emerges from this analysis is the existence of a geodesically complete space-time\footnote{Note that this is a purely gravitational effect of our framework and not of the matter sector, since the GR counterpart with Born-Infeld electrodynamics is geodesically incomplete \cite{Breton}.} divided in two sectors (one on each side of the wormhole throat) which are causally disconnected from each other and, therefore, cannot exchange information. This is due to the fact that radial null geodesics, which are associated to light rays, take an infinite time to reach any of the wormhole mouths, which means that these wormholes lie on the future (or past) boundary of the space-time and, therefore, endless geodesics will never reach this future (or past) boundary. This implies the possibility of a geodesic traveling infinitely from past to future remaining in a finite spatial region (on one of the sides of the wormhole) and avoiding singular behaviour \cite{Senovilla}. It is worth noting that in other scenarios geodesic completeness can be achieved even if time-like and null geodesics effectively go through regions with curvature divergences. This is the case, for instance, of Palatini theories with more general curvature dependencies \cite{ors15a}  and also of wormholes in GR under the standard thin-shell approximation \cite{Visser}. In these cases, one might be concerned with the impact of curvature divergences on physical observers \cite{congruences}, as studied in  \cite{Tipler,Krolak,Nolan}, but that is an aspect that does not affect the model presented here.

\section{Conclusions} \label{sec:V}

The results obtained in this paper show that even simple extensions of General Relativity, such as quadratic $f(R)$ gravity, may provide useful insights into the resolution of space-time singularities in conventional four-dimensional classical  scenarios. The new gravitational dynamics yields a geometry in which point-like singularities are replaced by a wormhole structure, of area $4\pi r_c^2$, while recovering the GR geometry a few $r_c$ units away from the wormhole throat. The non-trivial topology of the wormhole  provides an explicit implementation of Wheeler's charge-without-charge mechanism in such a way that the electric flux per unit area at the wormhole throat remains a universal constant. Though curvature divergences arise at the
wormhole throat, we show that the resulting geometry is geodesically complete and, therefore, nonsingular. We point out that this result follows from a combination of gravitational and matter actions that satisfy all the energy conditions, and does not rely on any particular quantum gravity scheme: the singularity avoidance is realized in terms of a classical space-time geometry.

Together with the results obtained in other theories of gravity including higher-order corrections in the curvature
invariants \cite{ors15a}, it seems that the singularity avoidance is a generic and robust feature of Palatini gravities, resulting
from the emergence of a topological wormhole structure close to the center of the black hole. This property is not unique of four dimensional models and persists in higher dimensional
scenarios \cite{WHe,higher-dim}. We also note that the same models that remove the point-like singularity in black holes, also
produce nonsingular cosmologies where the Big Bang singularity is replaced by a cosmic bounce \cite{Barragan,oor}. These
results might have relevant implications for the weak cosmic censorship conjecture, which establishes that singularities
produced in the gravitational collapse should always be hidden behind an event horizon, so as not to be seen by distant
observers \cite{WCCC}. The validity of this conjecture, originally introduced to ensure the predictability of the laws of physics, is still controversial \cite{Joshi}. Here we point out that if new high-energy physics occurring in the process of gravitational collapse could be
described in terms of metric-affine or Palatini theories, like the one considered here, the avoidance of singularities [understood as geodesically complete space-times] would remove the need for such a conjecture. To offer a complete answer to this question is still premature, as current research in gravitational collapse in Palatini theories is rather scarce. In this sense, our results point out that the removal of space-time singularities is a non-perturbative phenomenon, which would imply the validity of GR up to scales very close to the center of black holes. Research along these lines is currently under way.

\section*{Acknowledgments}

The work of C.B. is supported by NSFC (Chinese agency) grants No.~11305038 and No.~U1531117, the Shanghai Municipal Education Commission grant for Innovative Programs No.~14ZZ001, the Thousand Young Talents Program, Fudan University, and the Alexander von Humboldt Foundation. G.J.O. is supported by a Ramon y Cajal contract, the Spanish grants FIS2014-57387-C3-1-P and FIS2011-29813-C02-02 from MINECO, the grants i-LINK0780 and i-COOPB20105 of the Spanish Research Council (CSIC), the Consolider Program CPANPHY-1205388, and the CNPq project No. 301137/2014-5 (Brazilian agency). D. R.-G. is funded by the Funda\c{c}\~ao para a Ci\^encia e a Tecnologia (FCT) postdoctoral fellowship No.~SFRH/BPD/102958/2014, the FCT research grant UID/FIS/04434/2013, and the NSFC grant No.~11450110403. A.C.-A. wishes to thank the Department of Physics at Fudan University, where part of this work was performed, for their hospitality.

\end{document}